\documentclass[conference]{IEEEtran}
\IEEEoverridecommandlockouts
\usepackage[a4paper,left=0.625in, right=0.625in, top=0.7in,bottom=1.07in]{geometry}
\usepackage[utf8]{inputenc}
\usepackage{cite}
\usepackage{amsmath,amssymb,amsfonts}
\usepackage{algorithmic}
\usepackage{graphicx}
\usepackage{textcomp}
\usepackage{url}
\usepackage[caption=false]{subfig}
\usepackage[justification=centering]{caption}
\usepackage[ruled,vlined]{algorithm2e}
\usepackage{epsfig}
\usepackage{float}
\usepackage{color}
\usepackage{balance}
\usepackage{bbm}
\usepackage{textcomp}
\usepackage{enumitem}

\def\BibTeX{\text{B\kern-.05em{\sc i\kern-.025em b}\kern-.08em
    T\kern-.1667em\lower.7ex\hbox{E}\kern-.125emX}}

\newcommand{\beq}{\begin{equation}}
\newcommand{\eeq}{\end{equation}}

\floatstyle{ruled}
\newfloat{algorithm}{tbp}{loa}
\providecommand{\algorithmname}{Algorithm}
\floatname{algorithm}{\protect\algorithmname}

\def\BibTeX{{\rm B\kern-.05em{\sc i\kern-.025em b}\kern-.08em
    T\kern-.1667em\lower.7ex\hbox{E}\kern-.125emX}}
\begin{document}

\title{Energy-Efficient Dynamic Edge Computing with Electromagnetic Field Exposure Constraints \\
\thanks{This work has been partly funded by the European Commission through the H2020 project Hexa-X (Grant Agreement no. 101015956).}}

\author{Mattia Merluzzi, Serge Bories and Emilio Calvanese Strinati\\
CEA-Leti, Université Grenoble Alpes, F-38000 Grenoble, France\\
email:\{mattia.merluzzi, serge.bories, emilio.calvanese-strinati\}@cea.fr\vspace{-.3 cm}}

\maketitle

\begin{abstract}
We present a dynamic resource allocation strategy for energy-efficient and Electromagnetic Field (EMF) exposure aware computation offloading at the wireless network edge. The goal is to maximize the overall system sum-rate of offloaded data, under stability (i.e. finite end-to-end delay), EMF exposure and system power constraints. The latter comprises end devices for uplink transmission and a Mobile Edge Host (MEH) for computation. 
Our proposed method, based on Lyapunov stochastic optimization, is able to achieve this goal with theoretical guarantees on asymptotic optimality, 
without any prior knowledge of wireless channel statistics. Although a complex long-term optimization problem is formulated, a per-slot optimization based on instantaneous realizations is derived. Moreover, the solution of the instantaneous problem is provided with closed form expressions and fast iterative procedures. Besides the theoretical analysis, numerical results assess the performance of the proposed strategy in striking the best trade-off between offloading sum-rate, power consumption, EMF exposure, and E2E delay. To the best of our knowledge, this is the first work addressing the problem of energy and exposure aware computation offloading.
\end{abstract}


\section{Introduction}
The sixth generation of mobile communication networks (6G) targets a radical long-term transformation of wireless systems, with a pervasive deployment of computing resources at the edge, as well as a myriad of new connections spanning from conventional users to Internet of Things (IoT) and vertical sectors such as autonomous vehicles and Industry 4.0 \cite{CalvaneseGOWSC2021}. In this new perspective, human, physical and digital worlds will be embedded into the same ecosystem \cite{hexax}, making network management increasingly complex. In this context, extremely large amounts of capillary data will be wirelessly transferred between intelligent edge devices to operate data distillation, computation offloading services, training and inference of distributed (and/or federated) Artificial Intelligence (AI) and Machine Learning (ML) models, etc. In this direction, Multi-access Edge Computing (MEC)\cite{Pham2019Survey} supports the network evolution toward 6G \cite{6Gstrinati}, by bringing data processing at the network edge, thus close to the end consumers. Among several services, MEC networks allow resource poor end devices to offload computational tasks to nearby processing units. This comes with the challenge of a communication-computation co-design of wireless networks \cite{Merluzzi2020URLLC}, since computation offloading services involve both transmission and computation delays and energy consumption. Several works addressed the problem of striking the best trade-off between energy consumption, service delay and/or application level performance (e.g. AI/ML training/inference accuracy) \cite{Mao2017,Chen2019,MerluzziEML2021}. However, these works do not consider Electromagnetic Field (EMF) human exposure, which will increase due to continuous communications between the myriad of sensors and wireless Access Points (APs) and a drastic uplink traffic growth~\cite{Oueis2016}. Therefore, sustainable operations of future wireless networks impose to jointly consider: i) \textit{Performance}; ii) \textit{Energy efficiency}; iii) \textit{EMF exposure awareness}. While all these aspects have been previously disjointly investigated, a \textit{holistic} view is lacking and  needed towards future networks deployment \textit{and operation} phases. Indeed, while traditional network optimization takes into account Key Performance Indicators (KPIs) such as data rate and latency, 6G networks will be designed to also strengthen less tangible indicators, termed as Key Value Indicators (KVIs) \cite{hexax}, such as the EMF exposure, with the usual monitoring of new recommendations \cite{ICNIRP}, fundamental to run operators' networks with the best performance while not exceeding exposure limits, and also for a smooth public acceptance of new technologies.\\ \textbf{Related works.} Previous works focused on EMF-aware networking and hardware perspectives \cite{Zhao2016,Thors2016,Chiaraviglio2021}. %The authors in \cite{Zhao2016} derive the maximum allowed transmit power, based on the power density property of the phased arrays of mobile devices at 15 and 28 GHz. In \cite{Thors2016}, the authors study the EMF exposure in the case of array antennas for user equipment. A simulation-based method to optimize a MIMO 5G network operating at 3.7 GHz, with respect to power consumption and EMF exposure, is presented in \cite{Mat2018}.
The authors of \cite{Chiaraviglio2021} present the problem of network planning, considering a weighted function of gNB installation costs and 5G service coverage as objective, while \cite{jimenez2021} analyzes the impact of more or less strict regulations on network performance, showing how stricter limitations degrade network performance. This straightforward yet important result motivates us to investigate trade-offs between network power consumption, EMF exposure, and performance of MEC-enabled services. %Besides network deployment, also the optimization of network operations can help targeting network performance while guaranteeing low and/or regulation compliant EMF exposure.
%In this paper, we focus on network operations optimization for dynamic computation offloading, rather than focusing on EMF general studies.  
 In \cite{Dedo2015EMF}, the authors study a load balancing scheme in heterogeneous networks to achieve target data rate and EMF exposure. The work has been carried out within the project LEXNET \cite{Tesanovic2014}.\\
\textbf{Our contribution.} We focus on dynamic computation offloading, with end devices continuously generating data to be uploaded to a Mobile Edge Host (MEH) and processed \cite{Mao2017,Chen2019,Merluzzi2020URLLC,MerluzziEML2021}. %While in this paper we keep the application general, other works focus on specific applications such as edge learning tasks \cite{MerluzziEML2021}. We design a dynamic queueing system involving uplink and computation as, e.g., in . 
Exploiting Lyapunov stochastic optimization, we design an online algorithm able to jointly allocate radio and computing resources, achieving the maximum system sum-rate under stability constraints, as well as system power constraints, and exposure constraints over intended areas. Our method does not require any prior knowledge of wireless channel statistics, but only the solution of simple per-slot problems that enjoy closed form expressions or fast iterative procedures. Also, it comes with theoretical guarantees on sum-rate asymptotic optimality, system stability, as well as power and EMF exposure guarantees. To the best of our knowledge, this is the first work with a joint view of offloading performance, system power consumption, and EMF exposure.
\section{System model and Key Performance/Values}\label{sec:system_model}
\begin{figure}[t]
\centering
        \includegraphics[width=.9\columnwidth]{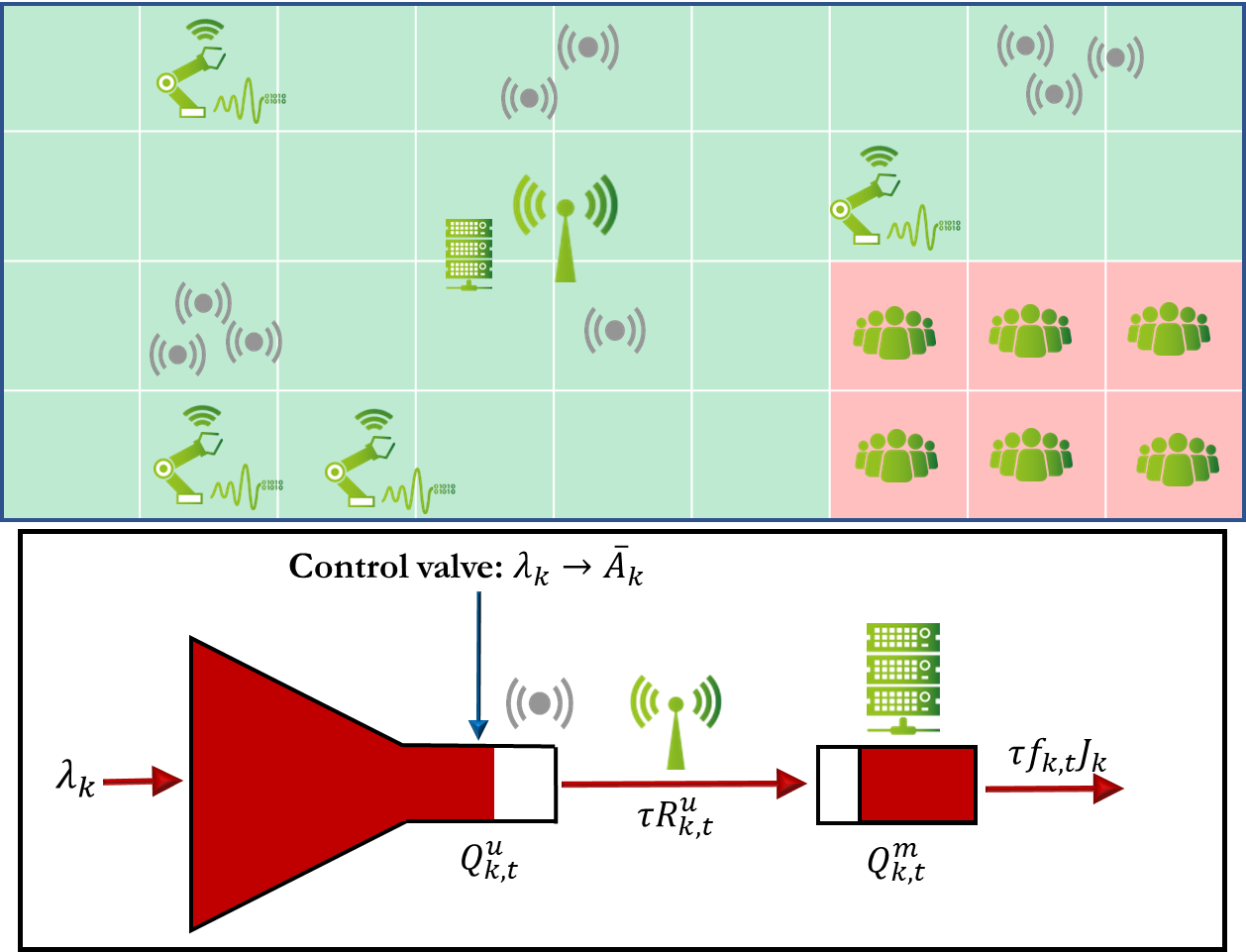}
    \caption{Scenario and arrivals}
    \label{fig:scenario_EMF}
  \vspace{-.4 cm}
\end{figure}
We consider a dynamic scenario, with time organized in slots $t=1,2,\ldots$ of duration $\tau$. In our setting, computation offloading entails two phases: i) Uplink data transmission; ii) Computation. Thus, we design a 2-hops queueing system, to model uplink transmission and computation. Let us define the long-term average of a random variable $X$ as 
\begin{equation}\label{average_value}
    \overline{X} = \lim_{T\to\infty}\frac{1}{T}\sum\nolimits_{t=1}^T\mathbb{E}\left\{X_t\right\},
\end{equation}
where the expectation is generally taken with respect to random context parameters (e.g. time-varying wireless channels). The above notation ($\overline{X}$) will be used throughout the paper to denote the long-term average of all involved variables. In each time slot $t$, every device $k=1,\ldots K$ stores $A_{k,t}$ new bits into an uplink buffer $Q_{k,t}^u$. As it will be clarified later on, $A_{k,t}$ will be optimized online to achieve the maximum sum-rate of offloaded data. At the same time, device $k$ possibly transmits data, buffered in previous slots, with data rate $R_{k,t}^u$. Thus, the uplink queue evolves as follows:
\begin{equation}\label{uplink_queue}
    Q_{k,t+1}^u=\max\left(0,Q_{k,t}^u-\tau R_{k,t}^u\right) +A_{k,t},\quad\forall k 
\end{equation}
The data transmitted by the end devices join a remote queue $Q_{k,t}^m, k=1,\ldots,K$, which is then drained by computing the tasks. Therefore, denoting by $J_k$ the number of bits computed by each CPU cycle, the queue evolution can be written as:
\begin{equation}\label{comp_queue}
    Q_{k,t+1}^m=\max\left(0,Q_{k,t}^m-\tau f_{k,t}J_k\right)+\min(\tau R_{k,t}^u,Q_{k,t}^u),
\end{equation}
where $f_{k,t}$ is the CPU frequency dedicated to device $k$ during slot $t$, which comes from the (optimized) MEH scheduling. 
\subsection{Average E2E delay}\label{sec:delay}
The simple queueing system described above is very useful to characterize the first two KPIs of our study: the average sum-rate and the E2E delay of computation offloading, measured as the time elapsed from the generation of new arrivals, until their computation at the MEH. Specifically, let us recall that a queueing system is stationary if all queues are stable, i.e. their long-term average is bounded. In particular, denoting by $\overline{Q_k^u}$, $\overline{Q_k^m}$ the long-term average of the involved queues (cf. \eqref{average_value}), the stability of the system is formalized as follows: 
\begin{align}\label{stability}
    & \overline{Q_k^u}<\infty,\;\forall k;\quad\overline{Q_k^m}<\infty,\;\forall k.
\end{align}
Moreover, if the system is stable, thanks to Little's law \cite{Little1961}, the average E2E delay is finite and can be written as follows:
\begin{equation}\label{e2edelay}
    \overline{D_k}=\tau(\overline{Q_k^u}+\overline{Q_k^m})/\overline{A_k},
\end{equation}
where $\overline{A_k}$ is the average number of new arrivals $A_{k,t}$ (cf. \eqref{uplink_queue}), computed as in \eqref{average_value}, to be optimized as we will clarify. %The average arrival rate represents our second KPI.
\subsection{Average data rate}\label{sec:rate}
Since a queueing system is stable if the average departure rate of each queue is greater than the average arrival rate, by designing a policy that stabilizes the system and, at the same time, maximize the average arrival rate, we guarantee the maximum system sum-rate \cite{Neely2008,Lakshminarayana2014}. Then, in this paper, our aim is to maximize the average sum-rate $\sum\nolimits_{k=1}^K\overline{A_k}$, to achieve the maximum sum-rate with finite E2E delay, and with long-term power and EMF exposure constraints as described later on. In particular, intuitively speaking, the idea is to admit, into the uplink queue, the maximum number of bits that the system is able to support (i.e. without instability), by limiting them through a fictitious \textit{control valve} that chokes the input data rate to match the system capacity \cite{Neely2008}, as depicted in Fig. \ref{fig:scenario_EMF}. For this purpose, to achieve stability, a \textit{joint} resource allocation policy for uplink transmission and MEH computation will also be designed, since, a priori, the system rate can be limited by the uplink, the processing capacity, or the EMF exposure limits to be guaranteed. The challenge is to devise a joint communication and computation resource management policy able to solve this problem without a priori knowledge of channel statistics.
\subsection{System power consumption}
In this paper, we consider the devices and the MEH as sources of power consumption, with the aim of keeping their average power expenditure under predefined thresholds. Given the uplink data rate $R_{k,t}^u$ a time $t$, the instantaneous power consumption of user $k$ can be written by inverting the well-known Shannon formula as 
$$\displaystyle p_{k,t}^u=\frac{N_0B_{k,t}^u}{h_{k,t}^u}\left(\exp\left(\frac{R_{k,t}^u\ln( 2)}{B_{k,t}^u}\right)-1\right),$$
where $B_{k,t}^u$ is the uplink bandwidth assigned to user $k$, $h_{k,t}^u$ is the time-varying uplink channel power gain, and $N_0$ is the noise power spectral density at the receiver. For simplicity, we assume a frequency division multiple access, without focusing on spectrum allocation. However, our method can be easily extended to incorporate bandwidth assignment. Finally, the MEH power consumption depends on the CPU core frequency $f_{s,t}$, which we assume to be dynamically scalable to save power \cite{Mao2017,Chen2019,Merluzzi2020URLLC}. Thus, the power consumption during time slot $t$ can be written as $p_{c,t}=\kappa f_{s,t}^3,$ where $\kappa$ is the effective switched capacitance of the processor \cite{Burd1996}. Note that, in this case, the constraint on the CPU should satisfy $\sum_{k=1}^K f_{k,t}\leq f_{s,t}$, where $f_{k,t}$ is defined in \eqref{comp_queue}, i.e. the sum overall all users cannot exceed $f_{s,t}$. Our aim, besides system stability, is to ensure that the long-term average of $p_{k,t}^u, \forall k$ and $p_{c,t}$ do not exceed predefined thresholds. We summarize the constraints as follows (cf. \eqref{average_value}):
\begin{equation}\label{long_term_power}
    \overline{p_k^u}\leq p_k^{\textrm{th}},\;\forall k,\quad\overline{p_c}\leq p_c^{\textrm{th}},
\end{equation}
where $p_k^{\textrm{th}}$ and $p_c^{\textrm{th}}$ are set a priori, based on the specific needs of devices (e.g. battery) and the MEC operator (e.g. OPEX).
\subsection{The EMF exposure as a Key Value Indicator}
As already mentioned, none of the previous works on dynamic computation offloading considered the EMF exposure. Inspired by \cite{Chiaraviglio2021}, where only communication aspects are tackled, let us consider a generic area populated by machines/sensors and humans. The area is divided in pixels as in the upper part of Fig. \ref{fig:scenario_EMF} and, in each pixel, we aim to guarantee a level of exposure below a predefined threshold. Moreover, the threshold can vary across space, due to specific needs or sensitive areas (e.g. with children), where the exposure has to be kept lower (e.g. over the red areas in Fig. \ref{fig:scenario_EMF}). To this aim, following ICNIRP recommendations \cite{ICNIRP}, we use the time-average incident power density, a measure that allows us to write the EMF exposure in closed form, given the channel power gains between devices and pixels. %Moreover, the ICNIRP recommendation on the average exposure over time naturally fits our methodology that guarantees a desired long-term behavior of the system. 
Since we consider the uplink exposure, let us write the total instantaneous power density at time $t$ over pixel $i$ as \cite{Chiaraviglio2021-2}
\begin{equation}\label{inst_pow_density}
    P_{d,t}^{i,\textrm{tot}}=\sum\nolimits_{k=1}^K\frac{4\pi}{\lambda^2} p_{k,t}^uh_{k,t}^i,\quad i\in\mathcal{P}
\end{equation}
with $\mathcal{P}$ the set of all pixels, $h_{k,t}^i$ is the instantaneous channel power gain between device $k$ and pixel $i$, and $\lambda$ is the wavelength. As already mentioned, our aim is to ensure that the long-term average of $P_{d,t}^{i,\textrm{tot}}$ does not exceed a threshold:
\begin{equation}\label{long_term_EMF}
    \overline{P_d^{\textrm{tot},i}}\leq P_{d}^{i,\textrm{th}},\quad \forall i\in\mathcal P
\end{equation}
As an example, ICNIRP recommendation for communications at $2$-$300$ GHz is $P_{d}^{i,\textrm{th}}=10$ W/m$^2$ for the general public, with $P_{d,t}^{i,\textrm{tot}}$ averaged over $30$ minutes. However, different values are used in national regulations or in specific conditions. The goal of this paper is to propose a method able to guarantee a predefined constraint $P_{d}^{i,\textrm{th}}$, rather than focusing on the values of $P_{d}^{i,\textrm{th}}$ over different scenarios. Then, we keep $P_{d}^{i,\textrm{th}}$ generic on purpose to specify it in the numerical results.
\section{Problem formulation and solution}\label{sec:problem}
Having defined and formalized all the KPIs and KVIs, let us now formulate the long-term optimization problem, aimed at maximizing the total average arrival rate, subject to stability constraints of all the queues involved in the offloading procedure, as well as exposure and power constraints. In this way, the long-term maximum sum-rate of the MEC system is attained \cite{Neely2008,Lakshminarayana2014}. The problem, is formulated as follows:
\begin{align}\label{problem_EMF}
    &\quad\underset{\{\mathbf{\Psi}_t\}_t}{\max}\quad\sum\nolimits_{k=1}^K\overline{A_k}\\
    &\textrm{subject to}\quad\eqref{stability},\;\eqref{long_term_power},\;\eqref{long_term_EMF},\quad(a)\: 0\leq A_{k,t}\leq A_k^{\max},\; \forall k,t;\nonumber\\
    &(b)\: 0\leq p_{k,t}^u\leq p_k^{\max},\; \forall k,t\quad(c)\: f_{s,t}\in\mathcal{F};\nonumber\\
    &(d)\: f_{k,t}\geq 0,\,\forall k,t\quad(e)\: \sum\nolimits_{k=1}^Kf_{k,t}\leq f_{s,t},\,\forall t,\nonumber
\end{align}
where $\mathbf{\Psi}_t=[\{A_{k,t}\}_{k=1}^K,\{p_{k,t}^u\}_{k=1}^K,f_{s,t},\{f_{k,t}\}_{k=1}^K]$. Besides the long-term constraints already discussed in Section \ref{sec:system_model}, the instantaneous constraints in \eqref{problem_EMF} have the following meaning: $(a)$ The new arrivals to the uplink queues are non negative and not greater a maximum value, to limit congestion \cite{Neely2008,Lakshminarayana2014}; $(b)$ The device uplink transmit power is non negative and lower than a  maximum value; $(c)$ The MEH CPU cycle frequency is chosen from a discrete set $\mathcal{F}$; $(d)$ The CPU frequency assigned to each user is non negative; $(e)$ The sum of the CPU frequencies assigned to each user does not exceed the selected MEH CPU cycle frequency. In principle, problem \eqref{problem_EMF} would require the knowledge of the uplink channel statistics to be solve optimally, thus making it very challenging. Moreover, even in case of perfect knowledge of the statistics, the problem is difficult to solve due to its non-convexity and the discrete nature of $f_{s,t}$ over a long-term horizon. 
However, we propose a simple solution based on Lyapunov stochastic optimization, inspired by \cite{Neely2008,Neely10,Lakshminarayana2014}. We will show how \eqref{problem_EMF} is first reduced to a pure stability problem that can be solved in a per-slot fashion, without requiring knowledge of the channel statistics. Furthermore, the per-slot problem is decoupled into radio and computation resource allocation problems that admit closed form solutions or fast iterative algorithms. Based on the theoretical foundations in \cite{Neely2008,Neely10,Lakshminarayana2014}, this solution is asymptotically optimal with respect to the original problem \eqref{problem_EMF} (i.e. the joint communication and computation co-design), 
with a single hyperparameter tuned to explore the trade-off between system sum-rate, E2E delay, power consumption, and EMF exposure. The reason for using Lyapunov tools to solve \eqref{problem_EMF} is twofold: i) It comes with theoretical guarantees on system stability, constraint violations, and asymptotic optimality; ii) Being a framework that guarantees long-term averages, it naturally fits the definition and ICNIRP recommendations of EMF exposure, in terms of average incident power density over time, as well as device and MEH power constraints.
\subsection{The Lyapunov approach}\label{sec:Lyapunov}
%Lyapunov stochastic optimization is a powerful tool used to first transform long-term problems into pure stability programs, which are then solved in a per-slot fashion \cite{Neely10}.  
Let us now introduce the tools needed to solve the complex problem in \eqref{problem_EMF}, with low complexity. Our first aim is to transform \eqref{problem_EMF} into a pure stability problem. Therefore, we define three \textit{virtual queues} used to guarantee constraints \eqref{long_term_power} and \eqref{long_term_EMF}. In particular, for each devices $k$, we define a virtual queue $Y_{k,t}$ evolving as follows (cf. \eqref{long_term_power}):
\begin{equation}\label{Y_evolution}
    Y_{k,t+1}=\max\left(0,Y_{k,t}+\epsilon_y\left(p_{k,t}^u-p_k^{\textrm{th}}\right)\right),\;\forall k,
\end{equation}
where $\epsilon_y>0$ is a step size. The role of the virtual queue is simple and intuitive: since $Y_{k,t}$ grows if the power constraint is instantaneously violated, and decreases otherwise, thus capturing the behavior of the system in terms of constraint violations. Similarly, for the MEH power consumption (cf. \eqref{long_term_power}), we define a virtual queue $H_t$ evolving as
\begin{equation}\label{H_evolution}
    H_{t+1}=\max\left(0,H_t+\epsilon_h\left(p_{c,t}-p_c^{\textrm{th}}\right)\right),
\end{equation}
with $\epsilon_h>0$. Finally, for constraint \eqref{long_term_EMF} we define, for each pixel, a virtual queue $Z_{i,t}$, $i\in \mathcal{P}$ evolving as
\begin{equation}\label{Z_evolution}
    Z_{i,t+1}=\max(0,Z_{i,t}+\epsilon_z(P_{d,t}^{i,\textrm{tot}}-P_{d}^{i,\textrm{th}})),
\end{equation}
with $\epsilon_z>0$. From \cite{Neely10}, we know that the \textit{mean rate stability}\footnote{For a virtual queue $G_t$, it is defined as $\lim_{T\to\infty}\mathbb{E}\{G_T\}/T=0$} of a virtual queue guarantees to meet the corresponding constraint. For instance, the mean rate stability of $Z_{i,t}, \forall i\in \mathcal{P}$ would ensure the EMF exposure limit. Therefore, our problem is to guarantee physical and virtual queues' stability (cf.  \eqref{uplink_queue}, \eqref{comp_queue}, \eqref{stability}, \eqref{Y_evolution}, \eqref{H_evolution}, \eqref{Z_evolution}). To this aim, letting
 $\mathbf{\Theta}_t=[\{Q_{k,t}^u\}_k,\{Q_{k,t}^m\}_k,\{Y_{k,t}\}_k,H_t,\{Z_{i,t}\}_i]$, we define the \textit{Lyapunov function} as follows \cite{Neely10}:
\begin{align}\label{lyap_fun}
    \mathcal{L}(\mathbf{\Theta}_t)=\frac{1}{2}\bigg[\sum\nolimits_{k=1}^K&[(Q_{k,t}^u)^2+(Q_{k,t}^m)^2+Y_{k,t}^2]\nonumber\\
    &+H_t^2+\sum\nolimits_{i\in\mathcal{P}}Z_{i,t}^2\bigg],
\end{align}
which is a measure of the overall congestion state of the system in terms of both physical and virtual queues. Based on \eqref{lyap_fun}, we can define the \textit{drift-plus-penalty} (DPP) function:
\begin{equation}\label{dpp}
    \Delta_{p,t}=\mathbb{E}\left\{\mathcal{L}(\mathbf{\Theta}_{t+1})-\mathcal{L}(\mathbf{\Theta}_t)-V\sum_{k=1}^KA_{k,t}\bigg|\mathbf{\Theta}_t\right\}.
\end{equation}
The DPP in \eqref{dpp} represents the conditional expected change of the Lyapunov function over one slot, with a penalty factor, weighted with parameter $V$, assigning more or less priority to the objective function of \eqref{problem_EMF} with respect to queue backlogs. Interestingly, if the DPP is upper bounded by a finite constant for each time slot, the  stability of the physical and virtual queues is guaranteed, thus guaranteeing constraints \eqref{long_term_power}, \eqref{long_term_EMF}, and finite delay (cf. \eqref{e2edelay}). Then, to push the network towards low congestion states, we proceed by minimizing a suitable upper bound of \eqref{dpp} \cite{Neely10}, which in this case reads as
\begin{align}\label{DPP_upper}
    \Delta_{p,t}\leq&\zeta+\mathbb{E}\bigg\{\sum\nolimits_{k=1}^k\big[Q_{k,t}^u\left(A_{k,t}-\tau R_{k,t}^u\right)\nonumber\\
    &+Q_{k,t}^m\left(\tau R_{k,t}^u-\tau f_{k,t}J_k\right)+\epsilon_yY_{k,t}(p_{k,t}^u-p_k^{\textrm{th}})\big]\nonumber\\
    &+\sum\nolimits_{i\in\mathcal{P}}\epsilon_z Z_{i,t}\left(P_{d,t}^{i,\textrm{tot}}-P_{d}^{i,\textrm{th}}\right)+\epsilon_h H_t(p_{c,t}-p_c^{\textrm{th}})\nonumber\\
    &-V\sum\nolimits_{k=1}^K A_{k,t}\bigg|\mathbf{\Theta}_t\bigg\},
\end{align}
where $\zeta$ is a positive constant omitted due to the lack of space, as well as the derivations leading to \eqref{DPP_upper}, which directly follow from \cite{Neely10}, using the following two inequalities: For a generic physical queue evolving as $Q_{t+1}=\max(0,Q_{t}-b)+A$, we can write $Q_{t+1}^2-Q_t^2\leq A^2+b^2+2Q(A-b)$; for a generic virtual queue $G_{t+1}=\max(0,G_t+g_t-\overline{g})$, we can write $G_{t+1}^2-G_t^2\leq (g_t^{\max}-\overline{g})^2+2G_t(g_t-\overline{g})$, where $g_t^{\max}$ is the instantaneous finite maximum value attainable by $g_t$, bounded by hypothesis. In particular, instantaneous power consumption and EMF exposure are clearly bounded by a finite constant. As in \cite{Neely2008,Neely10,Lakshminarayana2014} and based on stochastic optimization arguments, minimizing \eqref{DPP_upper} in each time slot theoretically leads to system stability under the assumption of i.i.d channel realization over consecutive time slots. Moreover, the optimal solution of \eqref{problem_EMF} (i.e. the maximum system sum-rate) is asymptotically achieved as the Lyapunov parameter $V$ increases, with the cost of increased average delay and convergence time \cite{Neely10}. In particular, the strategy that minimizes \eqref{DPP_upper} in each time slot achieve a $[O(1/V),O(V)]$ performance-delay trade-off, i.e. the distance between the value obtained by the online strategy and the optimal value of \eqref{problem_EMF} decreases as $O(1/V)$, with a cost of $O(V)$ increased upper bound on queue backlogs (i.e. delay and convergence time). More details follow from the theoretical results in \cite{Neely2008,Neely10,Lakshminarayana2014}, and are omitted due to the lack of space. Now, let us notice that, once the problem is cast in a per slot fashion, the flow control (i.e. data arrivals), uplink radio resource allocation, and CPU scheduling at the MEH, are decoupled into three independent sub-problems, which we present in the following, along with their respective solutions. Due to the lack of space, the derivations of the solutions will be summarized, but mainly come from simple mathematical manipulations involving classical convex optimization arguments that we will briefly recall \cite{boyd2004convex}. Note that the solutions of the following problems require the instantaneous knowledge of wireless channels, as well as physical and virtual queues' states. We assume them to be centrally solved by the MEH, although distributed implementations that go beyond the scope of this paper can be envisioned.
\subsection{Flow control sub-problem}\label{sec:flow}
The flow control problem involves the data arrivals (i.e. $A_{k,t},\forall k$) and is formulated as follows at time slot $t$:
\begin{align}\label{flow_control_problem}
    \underset{\{A_{k,t}\}_{\forall k}}{\min}\;&\sum\nolimits_{k=1}^K\left(Q_{k,t}^u-V\right)A_{k,t}\quad\nonumber\\
    &\textrm{s.t.}\quad 0\leq A_{k,t}\leq A_k^{\max},
\end{align}
Problem \eqref{flow_control_problem} is linear and admits the following straightforward simple closed-form solution for the optimal value $A_{k,t}^*$: $$A_{k,t}^*=A_k^{\max}\cdot\mathbf{1}\{Q_{k,t}^u\leq V\}, \forall k$$
where $\mathbf{1}\{\cdot\}$ denotes the indicator function.
\subsection{Uplink Radio Resource Allocation sub-problem}\label{sec:uplink}
The uplink resource allocation problem aims at optimizing the transmit power of each device, and is formulated as
\begin{align}\label{uplink_problem}
    \underset{\{p_{k,t}^u\}_{\forall k}}{\min}\;&-\sum\nolimits_{k=1}^K\left(Q_{k,t}^u-Q_{k,t}^m\right)\tau B_{k,t}^u\log_2\left(1+\frac{h_{k,t}^up_{k,t}^u}{N_0B_{k,t}^u}\right)\nonumber\\
    &\hspace{-0.6 cm}+\sum\nolimits_{i\in\mathcal{P}}\epsilon_z Z_{i,t}\sum\nolimits_{k=1}^K\frac{4\pi }{\lambda^2}p_{k,t}^uh_{k,t}^i+\sum\nolimits_{k=1}^K \epsilon_y Y_{k,t}p_{k,t}^u\nonumber\\
    &\textrm{subject to}\quad0\leq p_{k,t}^u\leq p_k^{\max},\quad \forall k;
\end{align}
Assuming the bandwidth to be assigned, the above problem can be decoupled among different users. Once decoupled, for a generic user $k$ let us notice that, if $Q_{k,t}^u\leq Q_{k,t}^m$, all terms in \eqref{uplink_problem} are monotone non-decreasing functions of $p_{k,t}^u$. Therefore, the optimal solution for the devices having  $Q_{k,t}^u\leq Q_{k,t}^m$ is $p_{k,t}^{u,*}=0$. For all other devices, the problem is strictly convex. Therefore, the optimal solution can be found in closed-form by solving the Karush-Kuhn-Tucker (KKT) conditions \cite{boyd2004convex} (omitted due to the lack of space), and is:
\begin{equation}
    \displaystyle p_{k,t}^{u,*}=\left[\sigma_{k,t}^u-\frac{N_0B_{k,t}^u}{h_{k,t}^u}\right]_0^{p_k^{\max}},\quad \forall k:Q_{k,t}^u> Q_{k,t}^m,\nonumber
\end{equation}
where $\sigma_{k,t}^u=\displaystyle\frac{\tau\left(Q_{k,t}^u-Q_{k,t}^m\right)B_{k,t}^u}{\displaystyle\ln(2)\left(\sum\nolimits_{i\in\mathcal{P}}\frac{4\pi}{\lambda^2}\epsilon_z Z_{i,t} h_{k,t}^i+\epsilon_y Y_{k,t}\right)}$. \\ Besides its theoretical mathematical derivation, this solution is also intuitive: if the local queue of device $k$ grows compared to its computation queue, device $k$ increases the transmit power to drain it; at the same time, if the virtual queues $Z_{i,t}$ and/or $Y_{k,t}$ grow, the transmit power is reduced to meet the long-term power and exposure constraints. Finally, the transmit power increases as $h_{k,t}^u$ increases, to exploit good radio channel conditions when transmitting to efficiently drain local queues. Let us notice that, even though for simplicity we assumed the bandwidth to be assigned a priori, for all devices transmitting at time $t$, the objective function of \eqref{uplink_problem} is convex also on $B_{k,t}^u$, making it suitable to optimize the spectrum allocation under the same optimization framework.
\subsection{Computation Resource Allocation sub-problem}\label{sec:comp}
For the CPU scheduling at the MEH, from \eqref{DPP_upper}, the sub-problem is cast as follows:
\begin{align}\label{computation_problem}
    &\hspace{-1.8 cm}\underset{f_{s,t},\{f_{k,t}\}_k}{\min}\quad \epsilon_h H_t\kappa f_{s,t}^3-\sum\nolimits_{k=1}^KQ_{k,t}^m\tau f_{k,t}J_k\\
    \textrm{subject to}\quad&(a)\: f_{s,t}\in\mathcal{F};\quad(b)\:f_{k,t}\leq \min\left(f_{s,t},\frac{Q_{k,t}}{J_k\tau}\right)\nonumber\\
    &(c)\: f_{k,t}\geq 0,\,\forall k,t\quad(d)\: \sum\nolimits_{k=1}^Kf_{k,t}\leq f_{s,t},\,\forall t,\nonumber
\end{align}
where, for efficiency purposes, we added constraint $(b)$ to ensure that each user cannot be assigned a clock frequency higher than the one necessary to empty the queue.
Problem \eqref{computation_problem} is a mixed integer non-linear program. However, For a fixed $f^s_{t}\in\mathcal{F}$, it is linear in $\{f_{k,t}\}_{k=1}^K$, and the optimal frequencies can be found using a simple procedure (requiring at most $K$ steps) that iteratively assigns largest portions of $f^s_{t}$ to the devices with largest values of $Q_{k,t}^m$, as also proved in \cite{Merluzzi2020URLLC,MerluzziEML2021}. %The procedure is iterative due to the fact that $f_{s,t}$ could be enough to empty the queue of a device. 
In the case a device's queue is completely drained, the remaining devices are allocated the available frequency until no frequency is left or all devices are served. Then, letting $\{f_{k,t}^*(f_{s,t})\}$ be the optimal frequencies assigned to the devices for each $f^s_{t}\in\mathcal{F}$, the optimal $f^s_{t}$ is $${f}^{*}_{s,t}= \underset{{f_{s,t}\in\mathcal{F}}}{\arg\min} \;\;-\sum\nolimits_{k=1}^K Q_{k,t}^m\tau J_k f_{k,t}^*(f_{s,t})+ \epsilon_h H_t \kappa(f_{s,t})^3,$$
which involves $|\mathcal{F}|$ evaluations of the objective function in \eqref{computation_problem}.
Finally, the overall procedure for Energy-efficient EMF-aware computation offloading is summarized in Algorithm \ref{alg:emf_aware_co}.
\begin{algorithm}[t]
\SetAlgoLined
\textbf{Input:} $\{Y_{k,0}\}_{k=1}^K$, $\{Z_{i,0}\}_i$, $H_0$, $\{Q_{k,0}^{u}\}_{k=1}^K$, $\{Q_{k,0}^{m}\}_{k=1}^K$. \\
In each time slot, observe $Q_{k,t}^{u}$, $Q_{k,t}^{m}$, $Z_{i,t}$, $Y_{k,t}^{u}$, $\forall k,i$, $H_t$:\\
\textbf{S1.} Observe $h_{k,t}^u,\forall k$, and find $A_{k,t}^*$ and $p_{k,t}^{u,*}$ as in Sections \ref{sec:flow} and \ref{sec:uplink}, respectively;\\
\textbf{S2.} Find $f_s^*(t)$ and $\{f_k^*(t)\}_k$ as in Section \ref{sec:comp};\\
\textbf{S3.} Update all queues as in \eqref{uplink_queue}, \eqref{comp_queue}, \eqref{Y_evolution}, \eqref{H_evolution}, \eqref{Z_evolution}.
\caption{Energy efficient EMF-aware comp.  offloading}
\label{alg:emf_aware_co}
\end{algorithm}
\vspace{-.4 cm}
\section{Numerical results}\label{sec:num_res}
\begin{figure*}[htb!]
    \centering
    \subfloat[Avg. EMF exposure vs. avg. sum-rate]{
        \includegraphics[width=0.315\textwidth]{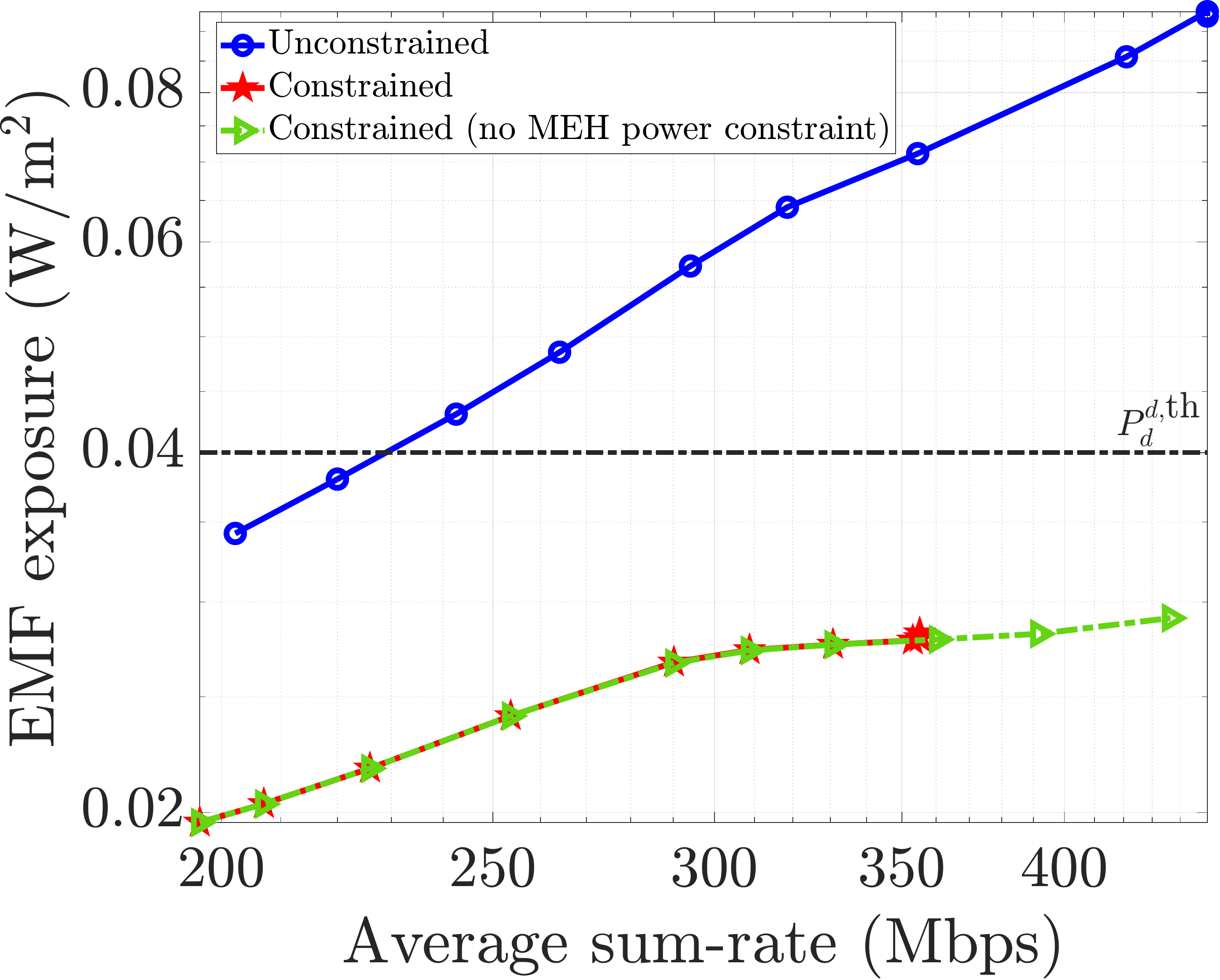}
        \label{fig:EMF_vs_rate}
    }
    \subfloat[Avg. sensor and MEH power vs. avg. sum-rate]{
        \includegraphics[width=0.32\textwidth]{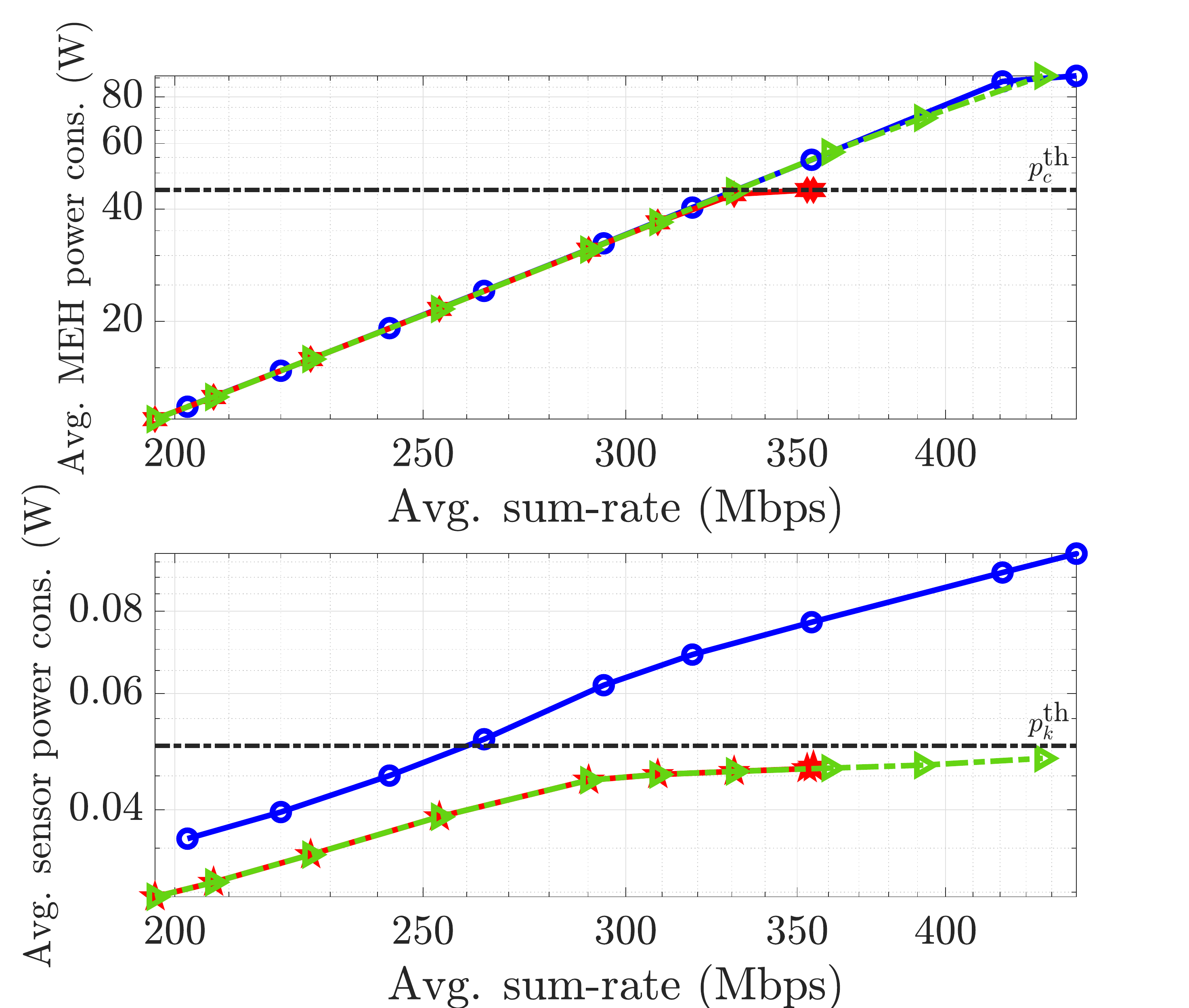}
        \label{fig:sensor_es_power_vs_rate}
    }
    \subfloat[Avg. E2E delay vs. avg. sum-rate]{
        \includegraphics[width=0.315\textwidth]{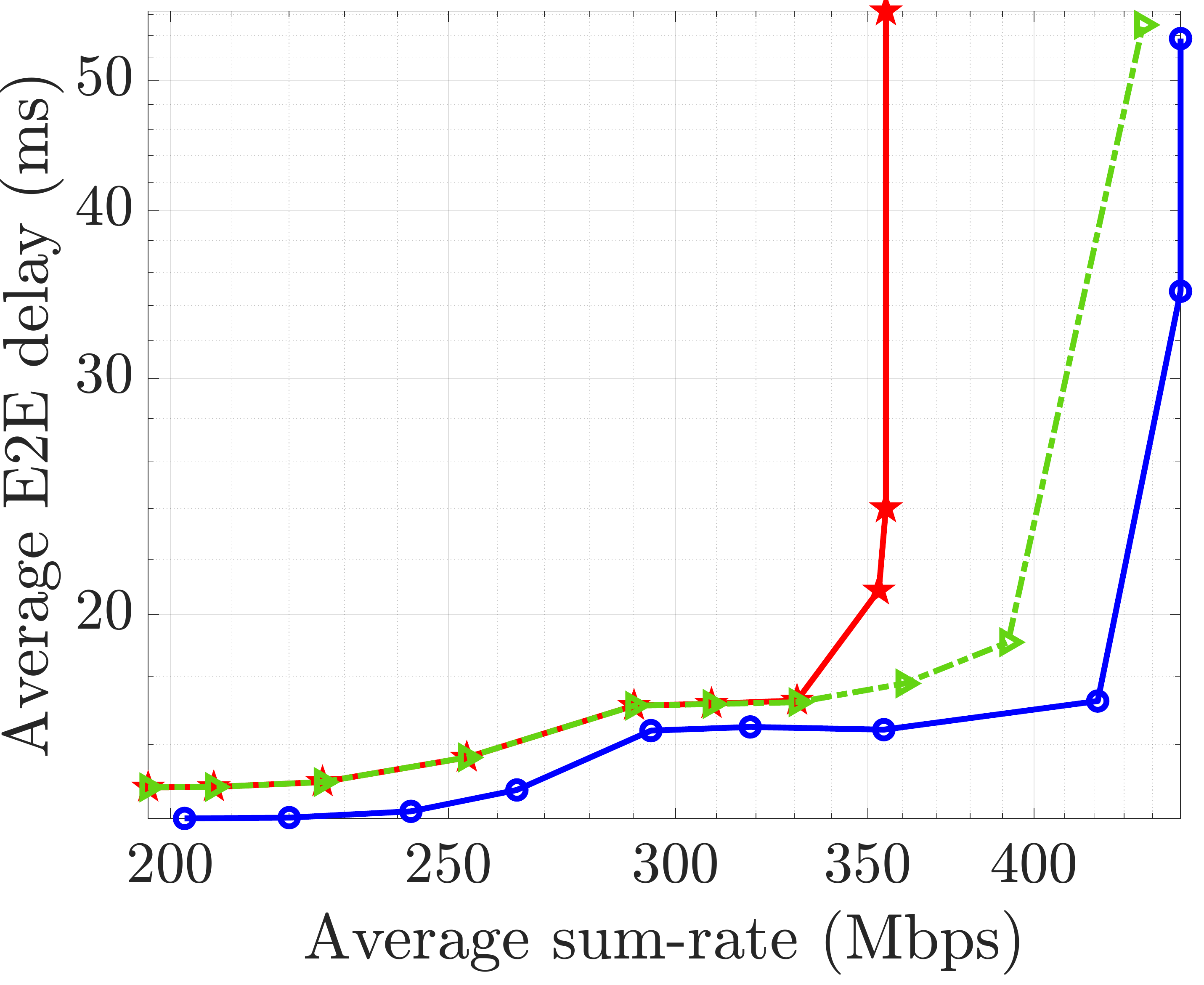}
        \label{fig:delay_vs_rate}
    }
    \caption{Trade-off between average sum-rate, EMF exposure, network power consumption and E2E delay}
    \label{fig:tradeoff}
 \vspace{-.4 cm}
\end{figure*}
For the numerical results, we consider an AP, located at the center of a squared area of side $15$ m, operating at $f_0=3.5$ GHz, with a total available bandwidth of $20$ MHz, equally shared among all devices. Around the AP, in the same area, we consider $100$ sensors uniformly randomly distributed in space, and transmitting with a  maximum power $p_k^{\max}=20$ dBm. Also, we assume a noise power spectral density $N_0=-174$ dBm/Hz. The path-loss model is taken from \cite{3gppTr382019}, considering a dense factory plant. A Rayleigh fading with unit variance and coherence time equal to the slot duration $\tau=5$ ms is assumed, while $A_k^{\max}=\tau R_{k}^{u,\max}$ (cf. constraint $(a)$ of \eqref{problem_EMF}). At the MEH side, the CPU clock frequency can be selected from the discrete set $\mathcal{F}=\{0,0.1,0.2,\ldots,1\}\times f_s^{\max}$, with $f_s^{\max}=4.5$ GHz. The effective switched capacitance of the processor is set to $\kappa=10^{-27}$. We use conversion factor $J_k=10^{-1}, \forall k$ (cf. \eqref{comp_queue}). The area around the AP is divided into $1$ m$^2$ pixels. All results are averaged for $2000$ slots and over $100$ different realizations of devices' positions.\\
\textit{\underline{Rate-exposure-energy-delay trade-off:}} As numerical results, we show the trade-off between sum-rate, EMF exposure, power consumption, and the E2E delay of the offloading service. We present three cases: i) \textbf{Unconstrained}: No limits on power and EMF are imposed, i.e. no virtual queues are used (cf. \eqref{Z_evolution}, \eqref{Y_evolution}, \eqref{H_evolution}), but only physical queues (cf. \eqref{uplink_queue}, \eqref{comp_queue}) for system stability. Therefore, this is the solution achieving the highest possible sum-rate, and represents our benchmark for the second setting; ii) \textbf{Constrained}: Constraints on EMF exposure over the space and power consumption (including sensors and MEH) are imposed, thanks to physical and virtual queues stability; iii) \textbf{Constrained (no MEH power constraint)}: Constraints on EMF exposure over the space and power consumption (including only sensors) are imposed.
Let us show the results in Fig. \ref{fig:tradeoff}, first with focus on the unconstrained setting, represented by the blue curve in all plots. In particular, Fig. \ref{fig:EMF_vs_rate} shows the EMF exposure as a function of the average sum-rate, obtained by increasing the Lyapunov parameter $V$ (cf. \eqref{dpp}) from left to right. As it is also clear from Fig. \ref{fig:delay_vs_rate}, we tuned $V$ to achieve similar E2E delay in the three cases, with a maximum value around $50$ ms. We can notice how, by increasing the average sum-rate, the EMF exposure increases as expected in the unconstrained setting, up to a maximum value around $90$ mW/m$^2$, corresponding to the maximum average sum-rate (around $450$ Mb/s). At the same time, in Fig. \ref{fig:sensor_es_power_vs_rate}, we show the average device and MEH power consumption as a function of the average sum-rate. %Here, the unconstrained setting is represented by the bourdeaux and the purple curves, for the ES and the devices, respectively. 
Again, by increasing the sum-rate, the average power consumption of both devices and MEH show an increasing behaviour towards its maximum value (100 mW and 90 W, respectively), corresponding again to the maximum average sum-rate on the abscissa. Finally, in Fig. \ref{fig:delay_vs_rate}, we show the average E2E delay (cf. \eqref{e2edelay}) as a function of the average sum-rate, showing its monotone increasing behaviour as expected by the theory ($[O(1/V)$,$(V)]$ trade-off). %Here, the unconstrained case is represented by the red curve.
Now that the unconstrained results have been described, let us focus on the constrained case, represented by the red curve in all plots. In particular, we impose: i) An EMF exposure threshold $P_d^{i,\textrm{th}}=40$ mW/m$^2$, represented by the black dashed horizontal line in Fig. \ref{fig:EMF_vs_rate}; ii) An MEH average power threshold $p_c^{\textrm{th}}=45$ W, and a device average power threshold $p_k^{\textrm{th}}=50$ mW, represented by the black dashed lines in Fig. \ref{fig:sensor_es_power_vs_rate}. First of all, we can notice how the online method is able to guarantee all the required constraints, in terms of exposure and power consumption, as also guaranteed by the theoretical arguments in Section \ref{sec:Lyapunov}. Moreover, EMF and sensor power consumption values are strictly lower than the thresholds, while the MEH power consumption approaches the desired threshold. This result suggests that the system sum-rate (around $350$ Mbps in this case) is limited by the MEH in this setting. To validate this conclusion, we run the same simulation, removing the MEH power constraint, thus achieving the result represented by the green curve in all plots. As expected, since the system is limited by the MEH capacity in the previous case, by removing this constraint, the method is able to guarantee sensor power consumption and EMF constraints, without degrading the performance in terms of sum-rate and E2E delay, as visible in Fig. \ref{fig:delay_vs_rate}. Therefore, we can conclude that our method is able to strike the best trade-off between average sum-rate, power consumption, EMF exposure and delay. Also, depending on the limiting capacity (computation or wireless), the method can dramatically reduce individual power consumption and/or EMF exposure, with negligible impact on network performance. More results involving different combinations of constraints are omitted due to the lack of space, but are foreseen for future investigations.
\section{Conclusions}
In this paper, we proposed an online algorithm to jointly optimize radio and computing resources for computation offloading purposes, to maximize the average sum-rate of offloaded data under EMF exposure and network power constraints, involving multiple end devices and an MEH. Our method does not require any prior knowledge of wireless channel statistics, and comes with theoretical guarantees on stability, exposure and power constraints, as well as on the asymptotic optimality of the solution in terms of average sum-rate. Besides the theoretical analysis, numerical results show the effectiveness of our method in striking the best trade-off between sum-rate, power consumption, EMF exposure, and service E2E delay. Future work includes multi-cell investigations with interference, directive communications, downlink communications and AP power consumption.  
\bibliographystyle{IEEEtran}
\bibliography{Mattia}

\end{document}